\documentclass[10pt,letterpaper]{article}
\usepackage{opex3}
\usepackage{cite}
\usepackage{amsmath}
\usepackage{float}

\begin{document}

\title{Modeling of the electromagnetic field and level populations in a waveguide amplifier: a multi-scale time problem}

\author{Alexandre Fafin,$^{*}$ Julien Cardin,$^{1}$ Christian Dufour, and Fabrice Gourbilleau}

\address{CIMAP, CNRS/CEA/ENSICAEN/UCBN \\ 6 boulevard Mar\'echal Juin, 14050 Caen cedex 4, France}

\email{$^{*}$alexandre.fafin@ensicaen.fr, $^{1}$julien.cardin@ensicaen.fr} 



\begin{abstract}A new algorithm based on auxiliary differential equation and finite difference time domain method (ADE-FDTD method) is presented to model a waveguide whose active layer is constituted of a silica matrix doped with rare-earth and silicon nanograins. The typical lifetime of rare-earth can be as large as some $\mathrm{ms}$, whereas the electromagnetic field in a visible range and near-infrared is characterized by a period of the order of $\mathrm{fs}$. Due to the large difference between these two characteristic times, the conventional ADE-FDTD method is not suited to treat such systems. A new algorithm is presented so that the steady state of rare earth and silicon nanograins electronic levels populations along with the electromagnetic field can be fully described. This algorithm is stable and applicable to a wide range of optical gain materials in which large differences of characteristic lifetimes are present.\end{abstract}



\section{Introduction}
The aim of this paper is to model the propagation of an electromagnetic field into an active optical waveguide. K. Yee in 1966\cite{Yee1966} presents the initial algorithm based on the finite difference time domain method (FDTD) used to discretize Maxwell's equations. in time and space so that it is suited to calculate the propagation of an electromagnetic field into dielectric media. In the past two decades, as computers have become more and more powerful, the FDTD method has met a growing success and has been extended to model antennas, periodic structures, dielectric materials exhibiting non linear dispersion \textit{etc}. \cite{Taflove1995}. One of the improvements to the basic FDTD method was to account for active dielectric materials which can absorb or emit the electromagnetic field. This improvement has been made by coupling Maxwell's equations to auxiliary differential eqations (ADE) describing the polarization densities linked to the authorized transitions and electronic level populations. 

For many years, rare earth ions have been used in silica-based optical amplifiers such as Erbium Doped Fibre Amplifier (EDFA)\cite{Miniscalco1991}. In these systems, the low gain value requires to employ significant length (10 to 15 $\mathrm{m}$) of doped fiber to achieve a workable power operation. In more compact system such as erbium-doped waveguide amplifier (EDWA) a higher gain has to be reach in order to shorten the operating length of the amplifier \cite{kik1998erbium}. One limiting factor of the gain is the low absorption cross section $\sigma_{abs}$ of rare earth ions. In order to increase this $\sigma_{abs}$, absorption sensitizers have been used such as ytterbium, semiconductor nanograins, metallic ions or organic complexes\cite{Polman2004}. Several studies have pointed out that silicon nanograins are efficient sensitizers and can increase by a factor of $10^{4}$ the effective absorption  cross section of rare earth ions \cite{Kenyon1994,Fujii1997}.  Erbium (Er$^{3+}$) has been the first rare earth studied due to the emission wavelength of 1.5 $\mathrm{\mu m} $, adapted to the telecommunications window in optical fibers\cite{Miniscalco1991}. However, there are three major gain limiting factors for the erbium ions: up-conversion, the excited state absorption and the re-absorption of the signal from the fundamental level. This last drawback is characteristic of a three levels system. More recently, neodymium ion has been proposed instead of erbium ion since its four levels configuration prevent signal re-absorption from the fundamental level.
 
Our goal is to model the propagation of an electromagnetic field into a waveguide with a layer containing absorbing and emitting centers as for example $\mathrm{Nd^{3+}}$ ions and silicon nanograins. More particularly, we want to determine the system characteristics as Fields, level populations, and gain in a steady state regime as a function of initial parameters such as concentration of emitting center, geometry, pumping configuration and pump and signal powers. Moreover, to model those steady states regime in a waveguide with a layer containing absorbing and emitting centers, we must take into account the time evolution of electromagnetic field and electronic levels populations of absorbing/emitting centers. In a waveguide containing silicon nanograins and neodymium ions, the typical lifetime of the electronic levels is about some $\mathrm{ms}$, whereas the characteristic period of the electromagnetic field is of the order of $\mathrm{fs}$. The choice of a common time step to treat such different time scales would require prohibitively long computation times (about $10^{15}$ iterations). One possible solution to overcome this multi-scale times issue was proposed in 2011\cite{Dufour2011}, by applying the so-called time scaling method which consists in multiplying the population rate differential equations by a scaling factor ($10^{6}$) so that the convergence of levels population is accelerated. Despite the number of required time iterations that has been reduced by $6$ order of magnitude, this technique does not sufficiently decrease the computation time and may lead to numerical instabilities for higher scaling factors.

In section \ref{oldalg}, we present the classical ADE-FDTD method and show that, within a reasonable computation time, the steady state of the system cannot be reached with such a difference between absorbing and emitting centers lifetimes and electromagnetic field period. Consequently, in section \ref{newalg}, we propose a new algorithm based on the ADE-FDTD method which allows to compute the electromagnetic field distribution, the gain, and the electronic levels populations in the waveguide in the steady state regime. Finally, we show in section \ref{results} the results of a calculation performed on a waveguide composed of silicon rich silicon oxide (SRSO) matrix containing silicon nanograins and neodymium ions.  

\section{Classical ADE-FDTD method}\label{oldalg}
The FDTD is based on time and space discretization scheme of Maxwell equations proposed by Yee \cite{Yee1966} which allows to calculate the propagation of  electromagnetic field ($\textbf{E}$,$\textbf{H}$) in time domain\cite{Taflove1995}. The ADE method consists in the use of extra terms such as current density $\textbf{J}$ or polarization density $\textbf{P}$ which are solutions of a differential equation with the aim to model some non linear optical behavior such as dispersive or gain media\cite{hagness1996subpicosecond}. The fields $\textbf{E}$, $\textbf{H}$ are treated by Maxwell equations rewritten as following:
\begin{equation}
		\left\{
		\begin{split}
		\nabla \wedge \textbf{E}&= -\mu \frac{\partial \textbf{H}}{\partial t} - \rho\textbf{H} \\
		\nabla \wedge \textbf{H}&= \epsilon_0\epsilon_r \frac{\partial \textbf{E}}{\partial t} + \frac{\partial \textbf{P}_{tot}}{\partial t}+ \sigma \textbf{E}
		\end{split}
		\right.	 
		\label{equations maxwell}
\end{equation} 

where $\epsilon_0\epsilon_r$ and $\mu$ are respectively the static permittivity and magnetic permeability. $\sigma$ is the usual electrical conductivity and $\rho$ is a fictitious magnetic resistivity used for boundary conditions of the calculation box. Berenger's perflectly matched layers (PML)\cite{Berenger1994} have been implemented as boundary conditions. Both $\sigma$ and $\rho$ have been chosen so that PML boundary conditions can minimize electromagnetic reflection and maximize absorption. $\textbf{P}_{tot}=\sum\textbf{P}_{ij}$ is the sum of all polarizations corresponding to each transition of the absorbing/emitting centers (hereafter: silicon nanograins and rare earth ions). The use of one polarization density $\textbf{P}_{ij}$ per optical transition between levels $i$ and $j$ allows the description of the global dynamic permittivity $\epsilon(\omega)$ of the matrix arising from the dipole moment densities induced by optical transitions in emitting centers.

The time and space steps must fulfill the classical stability conditions of FDTD calculation\cite{taflove1975numerical}:
\begin{itemize}
\item Space step $ \Delta < \frac{\lambda}{10} $, where $ \lambda $ is the smaller wavelength in the calculation. 
\item Time step $ \Delta t = S_{c} \frac{\Delta}{c \sqrt{d}} $, where $ c $ is the speed of light, $d=1,2,3$ depending on the dimensionality of the problem and $ S_c $ is the Courant number between 0 and 1 whose choice is empirical.
\end{itemize}

Neglecting the Rabi oscillation term\cite{chang2004finite}, for a transition between levels $i$ and $j$ the polarization density $\textbf{P}_{ij}$ is linked to the instantaneous electric field $\textbf{E}(t)$ and to the population difference $\Delta N_{ij}=N_i-N_j $ through the Lorentz type polarization density differential equation (\cite{Siegman1986}):
\begin{equation}
\frac{d^{2} \textbf{P}_{ij}(t)}{dt^2}+\Delta \omega_{ij} \frac{d \textbf{P}_{ij}(t)}{dt}+\omega_{ij}^2 \textbf{P}_{ij}(t) = \kappa_{ij} \Delta N_{ij} (t) \textbf{E}(t)
\label{polarisation}
\end{equation}
where $\Delta \omega_{ij}$ is the linewidth including radiative, non-radiative and dephasing processes of the transition\cite{Dufour2011}, and $\omega_{ij} $ is the resonance frequency of this transition. $\kappa_{ij}$ defined in \cite{Siegman1986} depends on the transition lifetime $\tau_{ij}$ and on the optical index $n$:  
\begin{equation}
	\kappa_{ij}= \frac{6 \pi \epsilon_0 c^3}{\omega_{ij}^2 \tau_{ij} n}
	\label{kappa}
\end{equation}

from Eq. (\ref{polarisation}) and the stability conditions in Lorentz media obtained by P. G. Petropoulos \cite{Petropoulos1994} another numerical stability condition appears:
\begin{equation}
\Delta t \leq \frac{2 \pi}{100\omega_{ij}}
\end{equation}

Finally, the time evolution of the electronic level populations $N_i$ is modelled by usual rate equations. For example in the case of a two level system, where $ N_1 $ is the fundamental level and $ N_2 $ the excited level, the rate equation. of the fundamental level is \cite{Nagra1998}: 
\begin{equation} 
\frac{dN_1 \left(t\right)}{dt}= - \frac{1}{\hbar\omega_{12}}\textbf{E}\left(t\right)\frac{d\textbf{P}_{21} \left(t\right)}{dt}+\frac{N_2\left(t\right)}{\tau_{21}\vert _{nr}^{r}}
\label{population}
\end{equation}

The term $\frac{1}{\hbar\omega_{12}}\textbf{E}\left(t\right)\frac{d\textbf{P}_{21} \left(t\right)}{dt}$ (in $\mathrm{ph.cm^{-3}.s^{-1}}$) is the induced radiation rate (resp. excitation rate) if it is negative (resp. positive). The terms N$_i$ (i=1,2) (in $\mathrm{cm^{-3}}$) are the population densities of different atomic levels, and $\tau_{21}\vert _{nr}^{r}$corresponds to the lifetime of spontaneous emission from the level 2 to the level 1. In principle, Eqs. (\ref{equations maxwell}), (\ref{polarisation}) and (\ref{population}) must be solved simultaneously. As explained above, in the visible and near infrared spectra the electromagnetic field has a characteristic time of the order of $10^{-15}~\mathrm{s}$. Furthermore, the excited levels have characteristic lifetimes as long as a few $\mathrm{ms}$\cite{Pacifici2003,Lee2005,Toccafondo2008}. Accordingly, due to the time step imposed by the ADE-FDTD method lower than $10^{-17}~\mathrm{s}$\cite{Oubre2004,Biallo2006}, the number of iteration must be as huge as $10^{15}$ in order to reach the steady states of the levels populations. So, a conventional calculation with the classical ADE-FDTD method where the equations of populations are calculated at the same time of the electromagnetic field is impossible in reasonable time.

\section{Development of the new algorithm}\label{newalg}
\begin{figure}[h!]
	\centering
	\includegraphics[width=0.9\textwidth, trim=5.5cm 15cm 5.5cm 2.75cm]{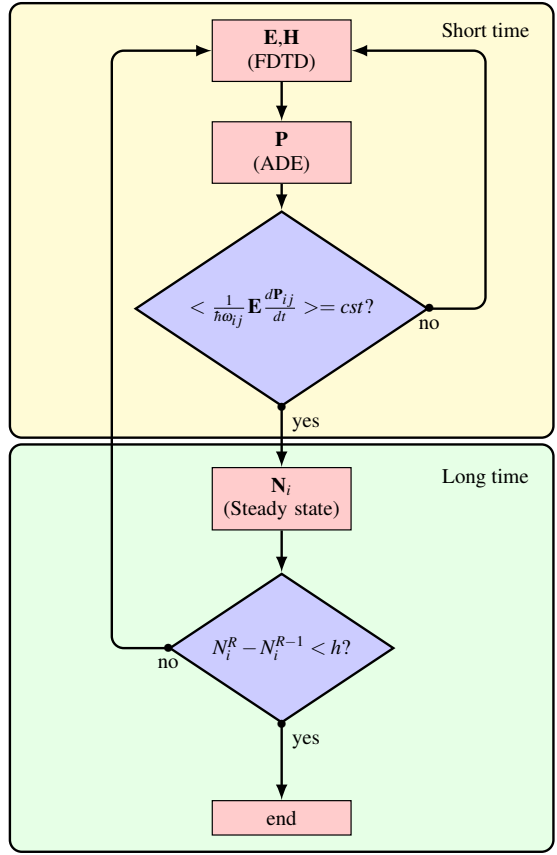}
	\caption{The new algorithm flowchart showing the alternation of the short time loop calculating electromagnetic field and polarizations and the long time loop including calculation of levels populations.}
	\label{algorithm}
\end{figure}

In order to reduce the computational time, we developed a new algorithm based on the ADE-FDTD method diagrammed in Fig. \ref{algorithm}. The choice of the linewidth $\Delta \omega_{ij}$, its link to the absorption cross section of emitting center and its consequence on the calculation duration will also be discussed.

\subsection{Explanation of the new algorithm}

Considering the timescale difference between the fields \textbf{E, H}, $\mathbf{P}_{ij}$ on the one hand and populations $N_i$ on the other hand, we propose to decouple Eqs. (\ref{equations maxwell}), (\ref{polarisation}) and (\ref{population}) into two sets of equations, solved one after the other: (i) the electromagnetic field and polarization Eqs.  (\ref{equations maxwell}) and (\ref{polarisation})  and (ii) calculation of steady state populations (eq \ref{population}). By analyzing the time evolution of volumic density of photon $ I_{ij}(t)=\frac{1}{\hbar \omega_{ij}} \textbf{E}\frac{d\textbf{P}_{ij}}{dt} $ obtained with classical ADE-FDTD method, we notice that $ I_{ij}(t) $ is a "quickly variable" function with a "slowly variable" envelope which reaches a stationary value after $10^5$ iterations. Moreover, it has been noticed that the time evolution of  populations $N_i(t)$, in Eq. (\ref{population}), is governed by the "slowly variable" evolution of the volumic density of photon $I_{ij}$ envelope. Based on these two observations, we propose a new ADE-FDTD algorithm divided in short and long time loops:

\begin{itemize}
\item In the short time loop, electromagnetic fields and polarizations are calculated Eqs. (\ref{equations maxwell}) and (\ref{polarisation}) assuming that all the levels populations are constant. Thus, for each transition, $ \Delta N_{ij} $ in Eq. (\ref{polarisation}) are constant and the current average value of photon volumic density $<I_{ij}>(t)$ is calculated. This latter follows the temporal evolution of the "slowly variable" envelope of  $I_{ij}(t)$. We exit from this short time loop of the algorithm when a stationary value $<I_{ij}>^{stat}$ is reached. This occurs when the relative difference between successive iterations becomes lower than a threshold value $\eta$
\begin{equation}
 \frac{<I_{ij}>(t^{n+1})-<I_{ij}>(t^{n-1})}{<I_{ij}>(t^n)} < \eta
\end{equation}
where $t^{n-1}$, $t^{n}$, and $t^{n+1}$ are three successive maxima as shown in Fig. \ref{flux}.
\begin{equation}
<I_{ij}>(t)=\frac{1}{t}\int_0^t I_{ij}(t^\prime) dt^\prime
\end{equation}
\begin{figure}[!h]
	\centering
	\includegraphics[width=0.8\textwidth,trim=1cm 0.25cm 1cm 2cm]{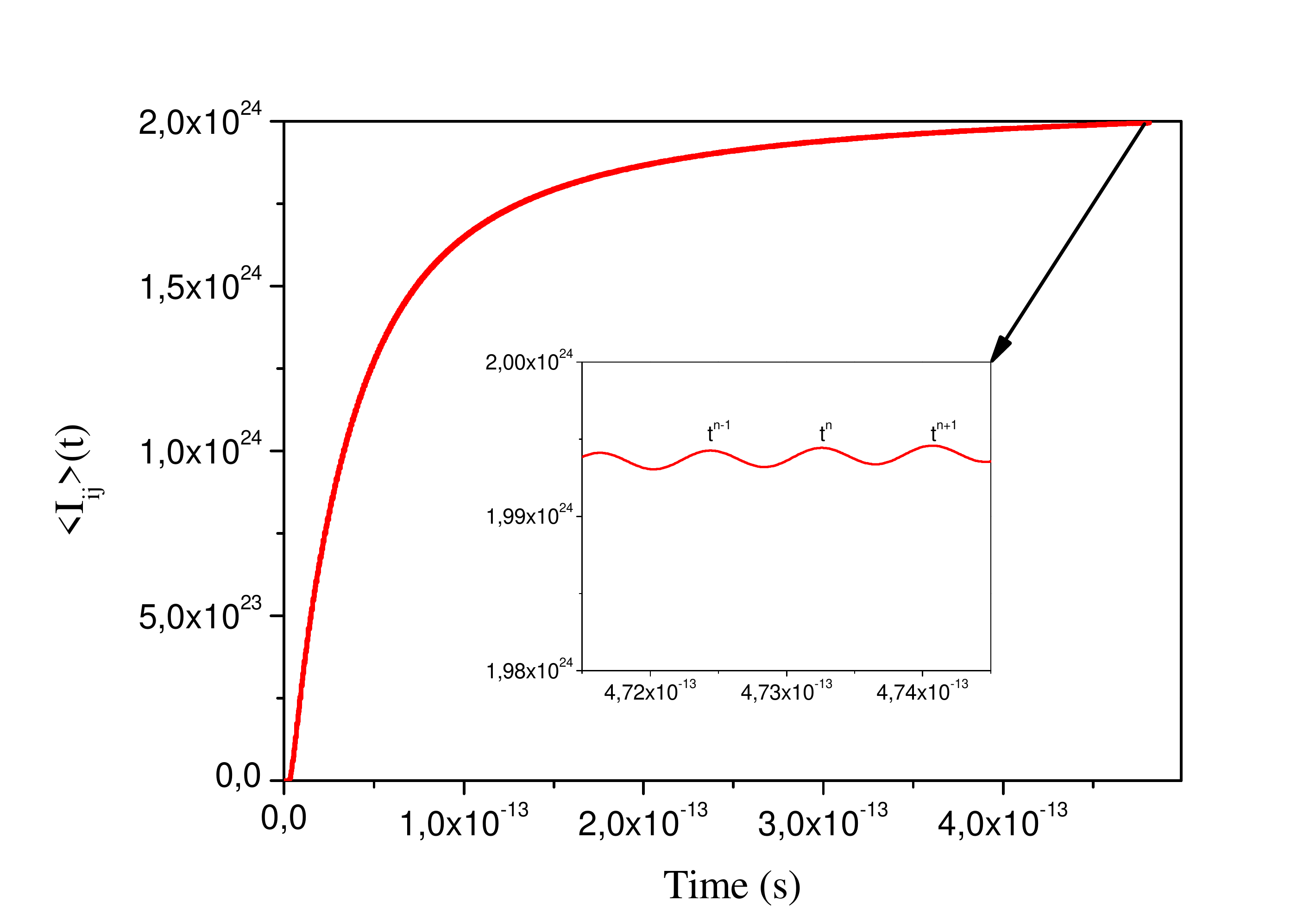}
	\caption{The typical evolution of $<I_{ij}>(t)$ when the populations do not vary and $\Delta N_{ij}>0$}
	\label{flux}
\end{figure}

\item In the long time loop of the algorithm, the levels population are calculated with Eq. (\ref{population}) by replacing the term  $\frac{1}{\hbar\omega_{12}}\textbf{E}\left(t\right)\frac{d\textbf{P}_{21} \left(t\right)}{dt}$ by its current average value in steady state $<I_{12}>^{steady}$ determined in the short time loop. More generally, referring to equations in sect. \ref{actlayer}, all  $\frac{1}{\hbar\omega_{ij}}\textbf{E}\left(t\right)\frac{d\textbf{P}_{ij} \left(t\right)}{dt}$ terms will be replaced by their current average value $<I_{ij}>^{steady}$. In the long time loop, levels population are calculated by seeking the analytic solution solutions of rate equations in steady states ($\frac{dN_i}{dt}=0$). Until the difference of levels population $N_i^R-N_i^{R-1}$ for all levels populations between two consecutive long time iterations $R-1$ and $R$ is greater than a threshold value $h$, we return to short time loop, otherwise the overall calculation is stopped.
\end{itemize} 
\begin{figure}[h!]
	\centering
	\includegraphics[width=0.7\textwidth,trim=1cm 0.25cm 1cm 0.4cm]{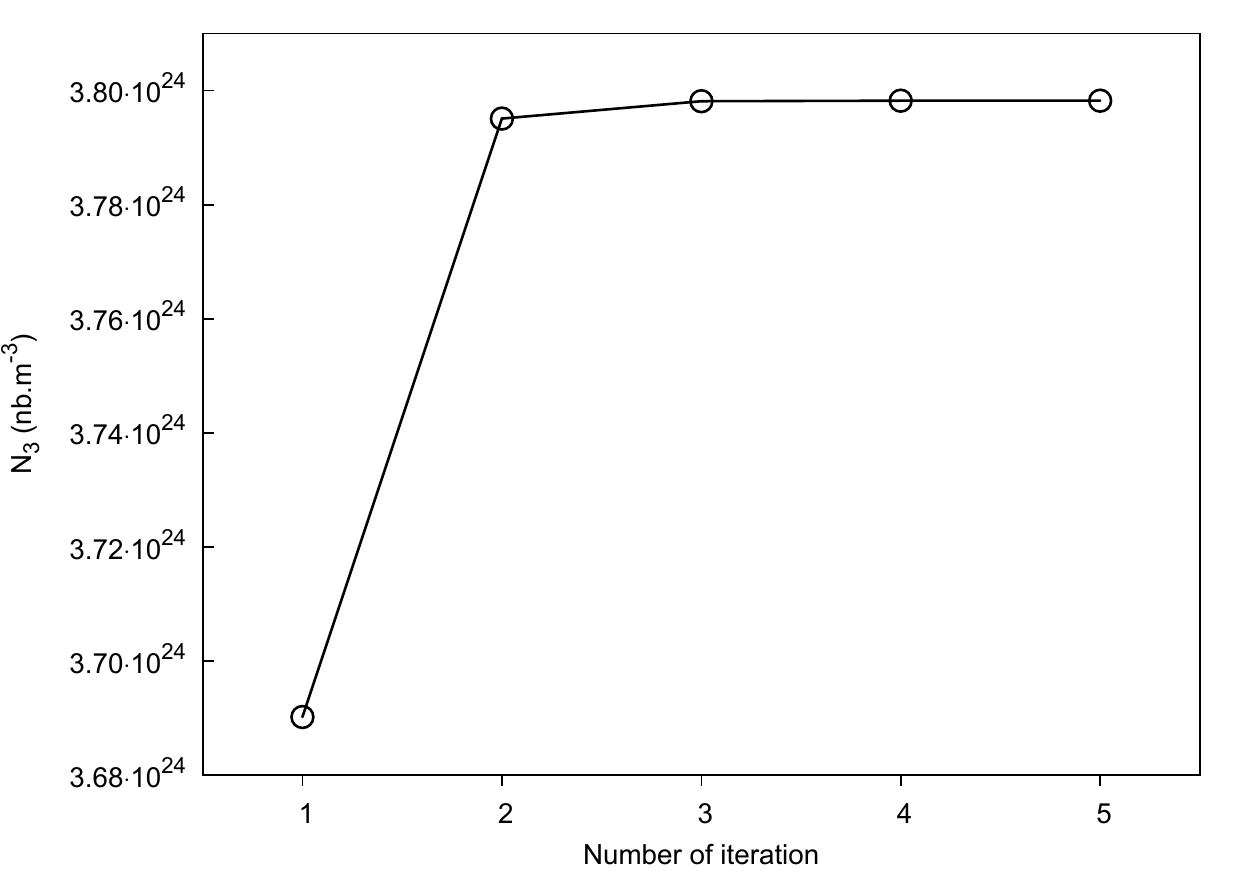}
	\caption{Evolution of the level $\mathrm{N_i}$ according to the number of long time iteration R}
	\label{pop}
\end{figure}
On Fig. \ref{pop} it can be noticed that the level population reaches a constant value in very few iterations. The overall number of iterations of the new algorithm is reduced from $10^{15}$ with the classical ADE-FDTD method to only $10^{5}$ resulting in a considerable reduction of calculation duration. The new algorithm is summarized in Fig. \ref{algorithm}.

\subsection{Choice of linewidth $ \Delta\omega $}\label{linewidth}
The description of an absorption or an emission process occurring during a transition $i$ to $j$ by the Lorentz oscillator polarization density Eq. (\ref{polarisation}) implies the equality of absorption and emission cross sections and that there is no inhomogeneous broadening. In order to calibrate the proper linewidth $\Delta\omega_{ij}$ of this Lorentz oscillator with respect to the absorption cross section, the Eq.  (\ref{polarisation}) is solved in forced harmonic regime resulting in a solution in the $\mathbf{P}=\epsilon_0 \left( \epsilon_r (\omega) -1 \right) \mathbf{E}$ form. This solution leads to the relationship between $\sigma(\omega)$ and $\Delta \omega_{ij}$:

\begin{equation}
\sigma(\omega)=\frac{\kappa_{ij} \omega_{ij}}{\epsilon_0 c} \left( \frac{\omega \Delta \omega_{ij}}{\left( \omega_{ij}^2 -\omega^2 \right)^2 + \omega^2 \Delta \omega_{ij}^2} \right)
\label{cross section 0}
\end{equation}

If the linewidth $\Delta \omega_{ij}$ is chosen at the resonance frequency $\omega=\omega_{ij}$, the Eq. (\ref{cross section 0}) is reduced to:

\begin{equation}
\sigma(\omega_{ij}) =\frac{\kappa_{ij}}{\epsilon_0 c} \frac{1}{\Delta \omega_{ij}}
\label{cross section}
\end{equation}

\begin{figure}[h!]
	\centering
	\includegraphics[width=0.8\textwidth]{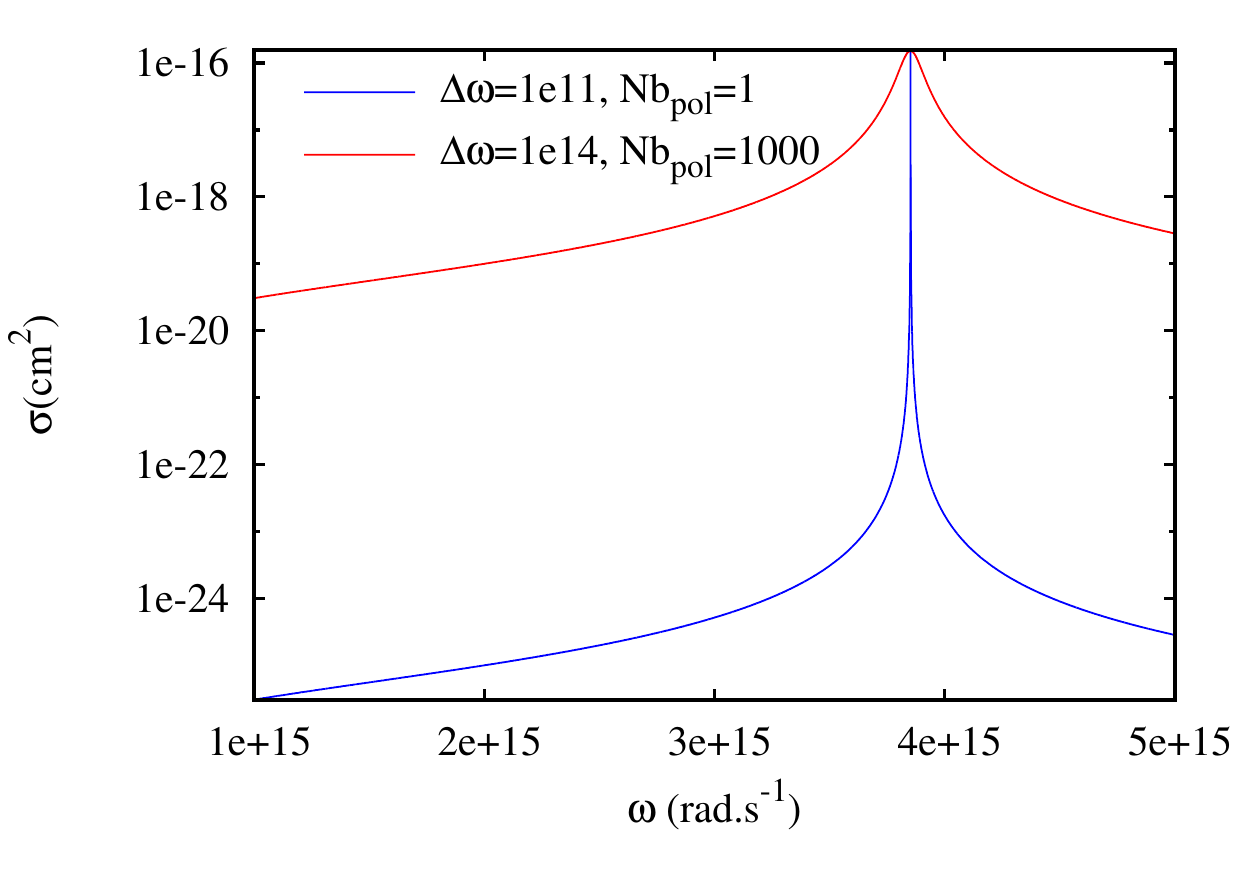}
	\caption{Cross section as a function of the pulsation with a transition at $ 3.8 \times 10^{15}~\mathrm{rad.s^{-1}}$}
	\label{section efficace}
\end{figure}

According to Eq. (\ref{cross section}), high absorption cross sections $\sigma$ (typically greater than $ 10^{-17}~\mathrm{cm^{2}} $) lead to small linewidths $\Delta \omega_{ij}$ (of the order of $10^{11}~\mathrm{rad.s^{-1}}$) which imposes a large number of iterations to reach a steady state. In order to calibrate the proper absorption cross section while keeping the number of iterations as small as possible, we exploit the superposition property of the polarization densities by using a number $N_p$ of identical polarization densities with larger $\Delta \omega_{ij}$. Despite, the fact that off-resonance cross sections ($\sigma(\omega)$ with $\omega \ne \omega_{ij}$) becomes wrong, its fast decreases off-resonance leads to a negligible effect. Fig. \ref{section efficace} shows the absorption cross sections as a function of $\omega$ for both cases: (i) $ \Delta \omega=10^{11}~\mathrm{rad.s^{-1}} $ with only one polarization ($N_P=1$), and (ii) $ \Delta \omega=10^{14}~\mathrm{rad.s^{-1}} $ with 1000 polarizations ($N_P=1000$). At resonance for $ \omega_{ij}=3.8 \times 10^{15}~\mathrm{rad.s^{-1}} $, the two methods lead to identical cross sections:$\sigma(\omega_{ij})_{(N_P=1)}=\sigma(\omega_{ij})_{(N_P=1000)}$. 

Eq. (\ref{cross section}) becomes into Eq. (\ref{cross section2}):\begin{equation}
\sigma = \frac{\kappa_{ij}}{\epsilon_0 c} \frac{1}{N_p \Delta \omega_{ij}}
\label{cross section2}
\end{equation}
With this method, the calibration of the cross section is made by choosing appropriate values of the number $ N_p $ of polarizations and linewidths $\Delta \omega_{ij}$.

\section{Results}\label{results}
\begin{figure}[!h]
	\centering
	\includegraphics[width=0.7\textwidth,trim=1cm 1cm 1cm 3cm,clip]{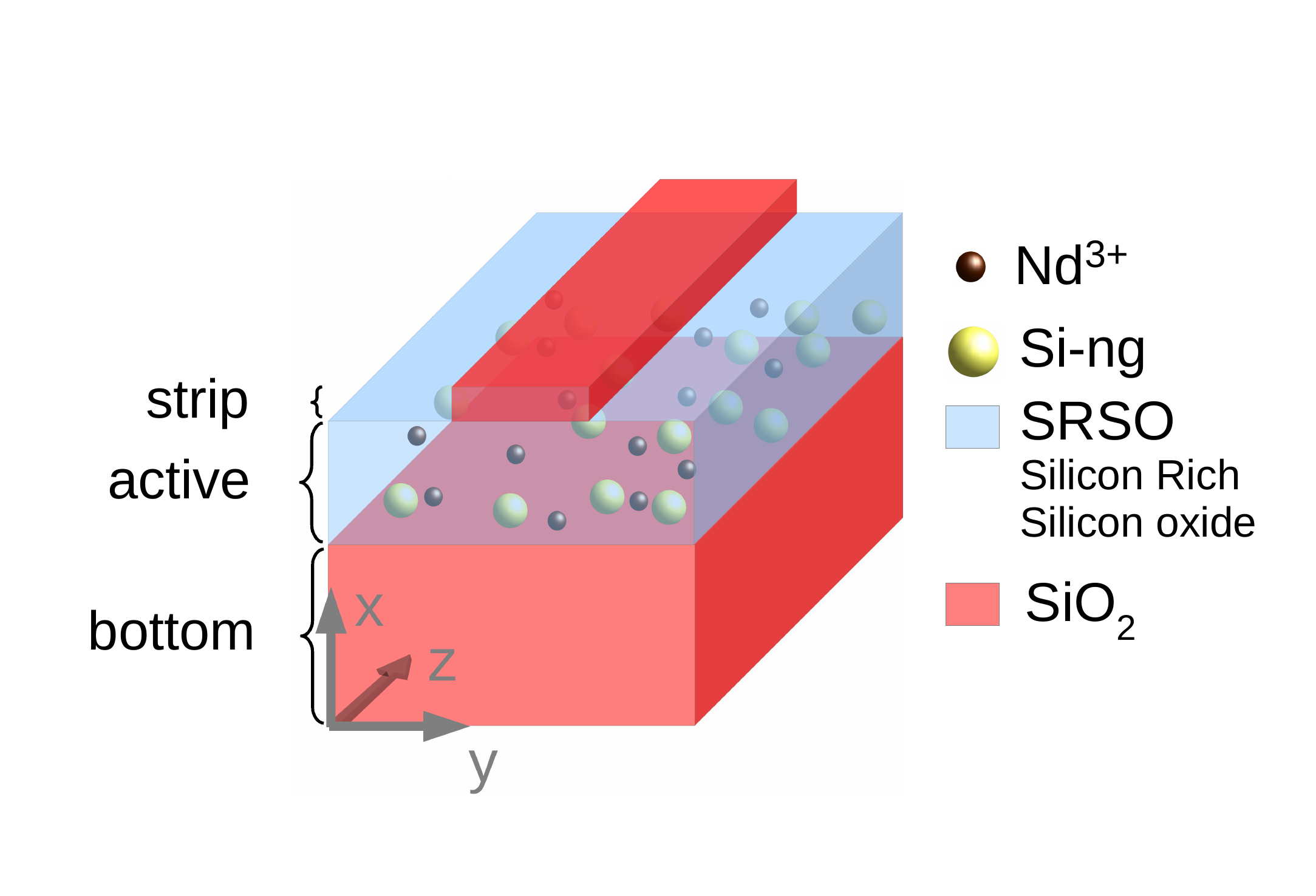}
	\caption{General view of the waveguide constituted by a bottom and strip cladding layers of silica surrounding the active layer constituted by silicon rich silicon oxide (SRSO) matrix doped with silicon nanograins (Si-ng) and Nd$^{3+}$ ions.}
	\label{waveguide}
\end{figure}

The present algorithm is applied to a waveguide composed of three layers as shown in Fig. \ref{waveguide}. Finite difference frequency domain method (FDFD) proposed by Fallakhair et al \cite{Fallahkhair2008} has been used to compute the electromagnetic mode profile. This eigenvalue problem of large sparse matrix has been solved using Fortran MUMPS library\cite{MUMPS:1,MUMPS:2}. The dimensions of the waveguide have been investigated so that the waveguide is monomode at the signal wavelength whereas obviously it is multimode at the pumping wavelength. We choose to inject the fundamental transverse electric (TE) mode related to the pumping field, in accordance with the experimental conditions. Moreover the FDTD algorithm was set up with the parameters reported in Table \ref{FDTDparam}.
\begin{table}[h!]
	\footnotesize
	\centering
	\caption{FDTD algorithm parameters}
	\begin{tabular}{ccccccc}
	\hline
	Parameter & $\Delta$ & $\Delta_t$ & $S_c$ & box length x& box length y& box length z\\
	\hline	
	Value   &$50$ nm & $10^{-17}$ s& $1$ & $169\Delta$ &$321\Delta$ &$161\Delta$\\
	\hline
	\end{tabular}
	\label{FDTDparam}
\end{table}

With these parameters, the maximum physical phase velocity error is -1.68\% and the maximum velocity-anisotropy error is 0.847\%\cite{Taflove1995}. Since we propagate single modes at the same wavelength and the wavefront distortion is small with respect to the wavefront of modes, we assume that this numerical dispersion is negligible in the results and conclusion that we will present in the paper.

\subsection{Description of the waveguide}	

The bottom cladding layer is composed of pure silica with a thickness of $ 3.5~\mathrm{\mu m}$. The active layer constituted of Silicon Rich Silicon Oxide (SRSO) contains silicon nanograins (Si-ng) and rare earth ions with a thickness of $ 2\mathrm{\mu m} $. A pure silica strip layer is stacked on the top of the SRSO layer. The width of the strip is $ 2~\mathrm{\mu m} $ and the thickness is $ 400~\mathrm{nm}$. The static refractive index of the active layer has been chosen greater ($1.5$) than the one of the strip and bottom cladding layers ($1.448$) to ensure the guiding conditions.

\subsection{Description of the active layer}\label{actlayer}
\begin{figure}[h!]
	\centering
	\includegraphics[width=0.7\textwidth,trim=0cm 0.5cm 0cm 0cm]{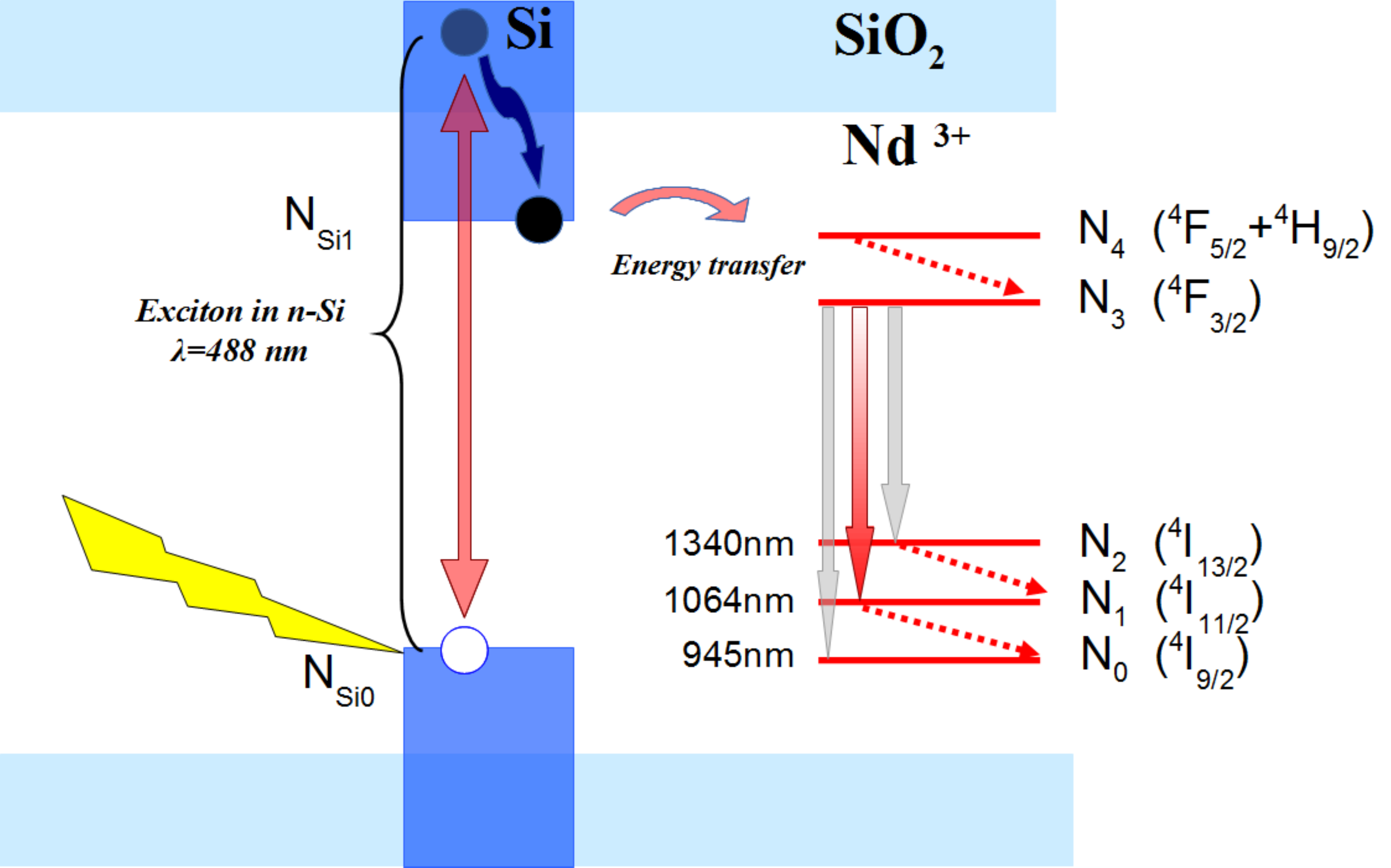}
	\caption{Excitation mechanism of rare earth}
	\label{excitation}
\end{figure}

The SRSO active layer contains silicon nanograins (Si-ng) and $ \mathrm{Nd^{3+}}$ that are modeled respectively by two levels and five levels systems as schematized in Fig. \ref{excitation}. 
The excitation mechanism of the $\mathrm{Nd^{3+}}$ ions is presented in Fig. \ref{excitation}. According to our experimental investigations\cite{Debieu2011} we pump the SRSO layer with an electromagnetic wave at $ 488~\mathrm{nm}$. The excitation mechanism leads to excitons generation in Si-ngs. Due to a low probability of multiexciton generation in a single Si-ng\cite{Govoni2012}, we assume a single exciton per Si-ng and therefore the Si-ng population will correspond to the exciton population. After non-radiative de-excitations, the exciton energy corresponds to the gap between the $ \mathrm{Nd^{3+}}$ levels $ \mathrm{N_0} $ ($^4 I_{9/2}$) and $ \mathrm{N_4} $ ($^4 F_{5/2} + ^4H_{9/2}$). Exciton can either transfer its energy to the $\mathrm{Nd^{3+}}$ ions  by dipole-dipole interaction and excites an electron from $\mathrm{N_0}$ to $\mathrm{N_4}$ or recombines radiatively or not to Si-ng ground level. After a fast non-radiative de-excitations from the level $\mathrm{N_4}$ to the level $\mathrm{N_3}$ ($^4 F_{3/2}$), we consider only the following three radiative transitions (we neglect the $^4 F_{3/2}\rightarrow ^4 I_{15/2}$ transition) : $N_3 \rightarrow N_0 (^4 F_{3/2}\rightarrow ^4 I_{9/2},\:\lambda_{30}=945nm)$, $ N_3 \rightarrow N_1 (^4 F_{3/2}\rightarrow ^4 I_{11/2},\:\lambda_{31}=1064nm)$ and $ N_3 \rightarrow N_2 (^4 F_{3/2}\rightarrow ^4 I_{13/2},\:\lambda_{32}=1340nm)$. 
De-excitations from level $ \mathrm{N_2} $ to level $ \mathrm{N_1} $ and from level $ \mathrm{N_1} $ to level $ \mathrm{N_0} $ are fast non-radiative transitions. Since the branching ratio of $\mathrm{N_3}\rightarrow \mathrm{N_1}$ transition is high ($> 50\%$) and the probability of re-absorption by $\mathrm{N_1}$ level is low, it is advantageous to use the $N_3 \rightarrow N_1 (^4 F_{3/2}\rightarrow ^4 I_{11/2})$ transition emitting at 1064 nm as a signal wavelength. We describe hereafter the full set of rate equations governing the levels populations.

\bigskip 

\paragraph*{\bf Silicon nanograins}
We consider a two level system  where $ \mathrm{N_{Si_0}} $ and $ \mathrm{N_{Si_1}} $ are respectively the ground level and excited level populations.
 
\begin{equation}
\frac{dN_{Si_1}\left(t\right)}{dt}=+\frac{1}{\hbar\omega_{Si_{10}}}\textbf{E}\left(t\right)\frac{d\textbf{P}_{Si_{10}} \left(t\right)}{dt}-\frac{N_{Si_1}\left(t\right)}{\tau_{Si_{10}}\vert _{nr}^{r}}-KN_{Si_1}\left(t\right)N_{0}\left(t\right)
\label{eq_si_1}
\end{equation}

\begin{equation}
\frac{dN_{Si_0}\left(t\right)}{dt}=-\frac{1}{\hbar\omega_{Si_{10}}} \textbf{E}\left(t\right)\frac{d\textbf{P}_{Si_{10}}\left(t\right)}{dt}+\frac{N_{Si_1}\left(t\right)}{\tau_{Si_{10}}\vert _{nr}^{r}}+KN_{Si_1}\left(t\right)N_{0}\left(t\right)
\label{eq_si_0}
\end{equation}
The term $ \frac{1}{\hbar\omega_{Si_{10}}}\textbf{E}\left(t\right) \frac{d\textbf{P}_{Si_{10}} \left(t\right)}{dt} $ is the photon density rate. The transfer coefficient between silicon nanograins and rare earth ions $K$ is chosen equal to $10^{-14}~\mathrm{cm^{3}.s^{-1}}$ according to Toccafondo et al.\cite{Toccafondo2008}. Based on Pacifici et al.\cite{Pacifici2003}, the decay time of the level $N_{Si_1}$ including radiative and non-radiative recombinations is fixed at $\tau_{Si_{10}}\vert _{nr}^{r}= 50~\mu s$. The Si-ng absorption cross section $\sigma_{Si}$ is taken equal to $10^{-16}~\mathrm{cm^{2}}$\cite{Priolo2001}. From Eq. (\ref{cross section2}) this leads to a linewidth of $ \Delta\omega_{Si}=1.5\times 10^{11}~\mathrm{rad.s^{-1}}$ with one polarization density($N_P=1$). According to the discussion in sect. \ref{linewidth}, in order to reduce the number of iterations, we choose a superposition of $ \mathrm{N_p=1000} $ polarization densities with a linewidth of $\Delta \omega_{Si}= 1.5\times 10^{14}~\mathrm{rad.s^{-1}}$ (see Table \ref{linewidthtable}). The Si-ng concentration is fixed at $ 10^{19}~\mathrm{cm^{-3}} $.

\paragraph*{\bf Neodymium ions}
The levels populations of $\mathrm{Nd^{3+}}$ are described by the following rate equations:
 
\begin{equation}
\frac{dN_{4}\left(t\right)}{dt}=-\frac{N_{4}\left(t\right)}{\tau_{43}\vert_{nr}}+KN_{Si_1}\left(t\right)N_{0}\left(t\right)
\label{eq_nd_4}
\end{equation}

\begin{equation}
\begin{split}
\frac{dN_{3}\left(t\right)}{dt}=&+\frac{1}{\hbar\omega_{30}}\textbf{E}\left(t\right)\frac{d\textbf{P}_{30}\left(t\right)}{dt}+\frac{1}{\hbar\omega_{31}}\textbf{E}\left(t\right)\frac{d\textbf{P}_{31}\left(t\right)}{dt}+\frac{1}{\hbar\omega_{32}}\textbf{E}\left(t\right)\frac{d\textbf{P}_{32}\left(t\right)}{dt} \\ &+\frac{N_{4}\left(t\right)}{\tau_{43}\vert_{nr}}-\frac{N_{3}\left(t\right)}{\tau_{30}\vert_{nr}^r}-\frac{N_{3}\left(t\right)}{\tau_{31}\vert_{nr}^r}-\frac{N_{3}\left(t\right)}{\tau_{32}\vert_{nr}^r}
\end{split}
\label{eq_nd_3}
\end{equation}

\begin{equation}
\frac{dN_{2}\left(t\right)}{dt}=-\frac{1}{\hbar\omega_{32}}\textbf{E}\left(t\right)\frac{d\textbf{P}_{32}\left(t\right)}{dt}+\frac{N_{3}\left(t\right)}{\tau_{32}\vert_{nr}^r}-\frac{N_{2}\left(t\right)}{\tau_{21}\vert_{nr}}
\label{eq_nd_2}
\end{equation}

\begin{equation}
\frac{dN_{1}\left(t\right)}{dt}=-\frac{1}{\hbar\omega_{31}}\textbf{E}\left(t\right)\frac{d\textbf{P}_{31}\left(t\right)}{dt}+\frac{N_{3}\left(t\right)}{\tau_{31}\vert_{nr}^r}-\frac{N_{1}\left(t\right)}{\tau_{10}\vert_{nr}}+\frac{N_{2}\left(t\right)}{\tau_{21}\vert_{nr}}
\label{eq_nd_1}
\end{equation}

\begin{equation}
\frac{dN_{0}\left(t\right)}{dt}=-\frac{1}{\hbar\omega_{30}}\textbf{E}\left(t\right)\frac{d\textbf{P}_{30}\left(t\right)}{dt}+\frac{N_{3}\left(t\right)}{\tau_{30}\vert_{nr}^r}+\frac{N_{1}\left(t\right)}{\tau_{10}\vert_{nr}}-KN_{Si_1}\left(t\right)N_{0}\left(t\right)
\label{eq_nd_0}
\end{equation}

Similar to the case of silicon nanograins, we calculate the linewidth $\Delta \omega_{ij} $ for the different Nd$^{3+}$ transitions by Eq. (\ref{cross section2}). Since in literature, depending of the host matrix\cite{Siegman1986}, the emission cross section of $\mathrm{Nd^{3+}} $ at 1064 nm varies from $3\times10^{-20}\mathrm{cm^2}$ to $4.6\times10^{-19}\mathrm{cm^2}$, we choose to test two extreme values of Nd$^{3+}$ ions emission cross section equal to $\sigma=10^{-19}~\mathrm{cm^{2}}$ and $\sigma=10^{-20}~\mathrm{cm^{2}}$. For each transition, the corresponding linewidth values are high enough with regard to the number of iterations to reach steady state with one polarization density ($\mathrm{N_P=1}$). All the parameters discussed here are gathered in tables \ref{linewidthtable} and \ref{lifetime}. The concentration of $\mathrm{Nd^{3+}}$ is fixed at $10^{19}~\mathrm{cm^3}$.

\begin{table}[h!]
	\centering
	\small
	\caption{Pulsation, linewidth and number of polarizations chosen of radiative transitions}
	\begin{tabular}{cccc}
	\hline
	Transition &  Pulsation ($\mathrm{rad.s^{-1}}$) & Linewidth ($\mathrm{rad.s^{-1}}$) & Nb of polarizations\\
	\hline
	$ N_{Si_{1}} \rightarrow N_{Si_{0}} $ & $ \omega_{Si_{10}}=3.86\times 10^{15} $ & $\Delta \omega_{Si}=1.5\times 10^{14} $ & 1000\\
	$ N_3 \rightarrow N_2 $  & $ \omega_{32}=1.34 \times 10^{15} $& $\Delta \omega_{32}=6.683\times 10^{15} $ & 1\\
	$ N_3 \rightarrow N_1 $  & $ \omega_{31}=1.77 \times 10^{15} $& $\Delta \omega_{31}=1.786\times 10^{15} $ ($ \sigma=10^{-20}~\mathrm{cm^2} $) & 1\\
$ N_3 \rightarrow N_1 $  & $ \omega_{31}=1.77 \times 10^{15} $ & $\Delta \omega_{31}=1.786\times 10^{14} $ ($ \sigma=10^{-19}~\mathrm{cm^2} $)& 1\\
	$ N_3 \rightarrow N_0 $  & $ \omega_{31}=1.99 \times 10^{15} $& $\Delta \omega_{30}=1.054\times 10^{15} $ & 1\\
	\hline
	\end{tabular}
	\label{linewidthtable}
\end{table}

\begin{table}[h!]
	\footnotesize
	\centering
	\caption{Lifetimes of different transitions for Si-ng and $ \mathrm{Nd^{3+}}$}
	\begin{tabular}{cccccccc}
	\hline
	Transition & $ N_{Si_{1}} \rightarrow N_{Si_{0}} $ & $ N_4 \rightarrow N_3 $ & $ N_3 \rightarrow N_2 $ & $ N_3 \rightarrow N_1 $ & $ N_3 \rightarrow N_0 $ & $ N_2 \rightarrow N_1 $ & $ N_1 \rightarrow N_0 $\\
	\hline	
	Lifetime ($\mu s$) & $ 50 $ & $230\times10^{-6}$& $1000$ & $200$ & $ 250$ & $970\times10^{-6}$ & $ 510\times10^{-6}$ \\
	\hline
	\end{tabular}
	\label{lifetime}
\end{table}

\subsection{Field map}
\begin{figure}[h!]
	\centering
	\includegraphics[width=0.49\textwidth,trim=0cm 0cm 0cm 1cm,clip]{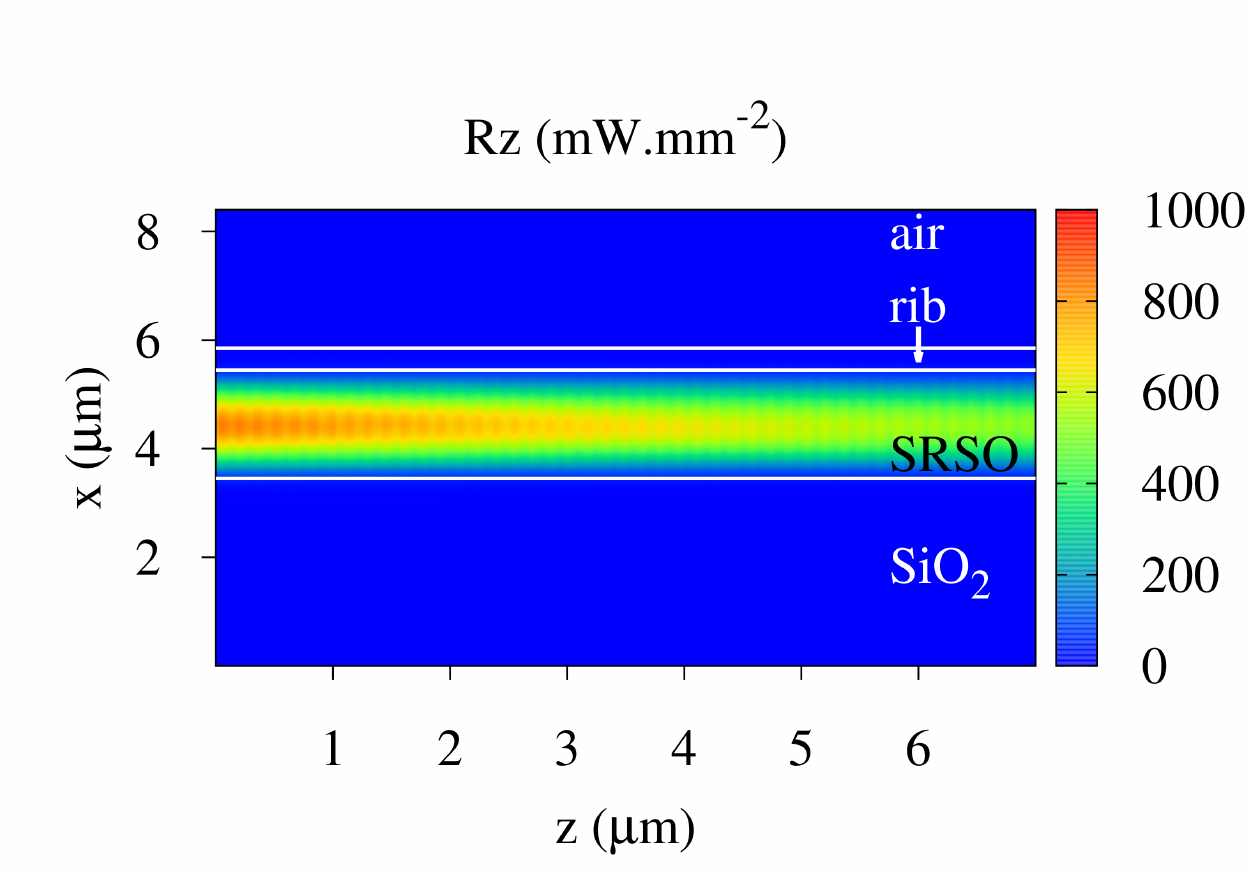}
	\includegraphics[width=0.49\textwidth,trim=0cm 0cm 0cm 1cm,clip]{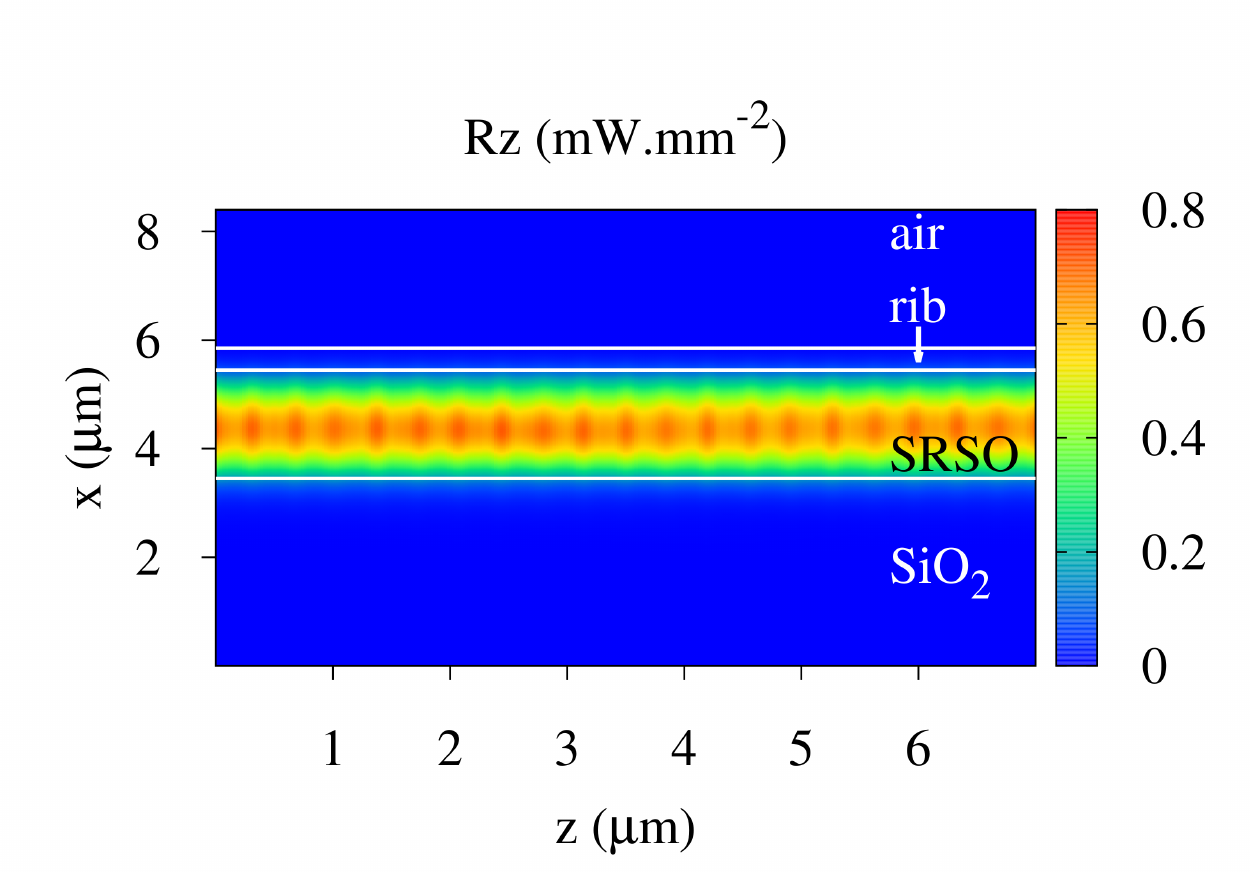}
	\caption{Longitudinal section view of Z-component of the Poynting vector of the pump (on the left) and of the signal (on the right). The pump ($ \lambda_s=488~\mathrm{nm} $) and the signal ($ \lambda_p=1064~\mathrm{nm} $) power injected are respectively $ 1~\mathrm{W.mm^{-2}} $ and $ 1~\mathrm{mW.mm^{-2}} $}
	\label{fieldmap}
\end{figure}

Poynting vector ($\textbf{R}=\textbf{E}\times\textbf{H}$) expressed in $\mathrm{W.mm^{-2}}$ has been calculated from electromagnetic field ($\textbf{E}$, $\textbf{H}$) compute by ADE-FDTD method.
After a time Fourier transform of $\textbf{R}(t)$, we determine the main z component of the pump ($488$nm) and signal ($1064$nm) intensities of $\textbf{R}$ in longitudinal section view [Fig. \ref{fieldmap}]. We can notice that, for both wavelengths, the electromagnetic field is well guided within the SRSO active layer of the waveguide. Due to the absorption at 488 nm by the silicon nanograins, a decrease of the pump intensity $R_{z,pump}$ along the waveguide is observed. However, the signal intensity $R_{z,signal}$ at 1064 nm does not appear absorbed over the 7$\mathrm{\mu m}$ length of the waveguide.

\begin{figure}[h!]
	\centering
	\includegraphics[width=0.49\textwidth,trim=0cm 0cm 0cm 1cm,clip]{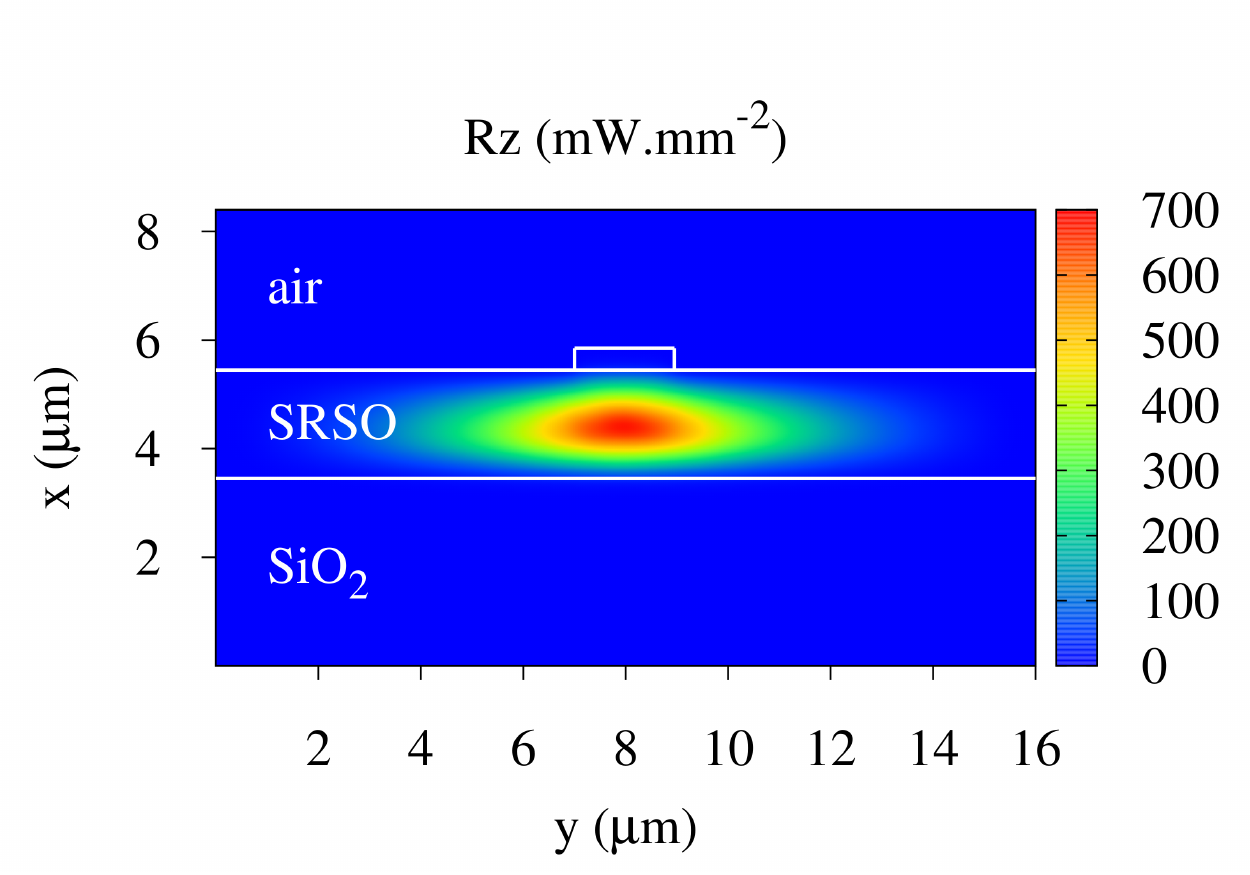}
	\includegraphics[width=0.49\textwidth,trim=0cm 0cm 0cm 1cm,clip]{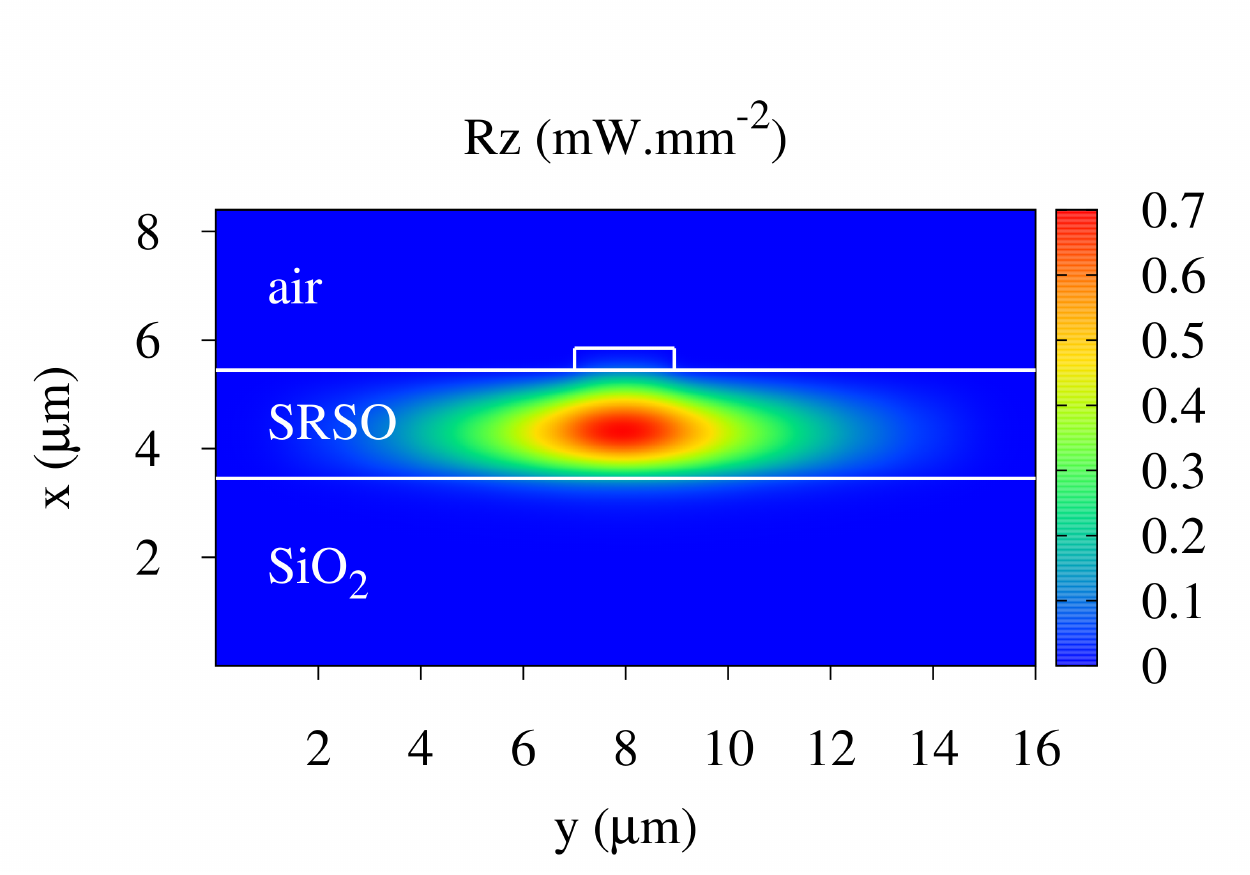}
	\caption{Transverse section view of Z-component of the Poynting vector of the pump (on the left) and of the signal (on the right) in the middle of the waveguide. The pump ($ \lambda_s=488~\mathrm{nm} $) and the signal ($ \lambda_p=1064~\mathrm{nm} $) power injected are respectively $ 1~\mathrm{W.mm^{-2}} $ and $ 1~\mathrm{mW.mm^{-2}} $}
	\label{fieldmap2}
\end{figure}

Figure \ref{fieldmap2} shows a transverse view of the $R_{z,pump}$ (left) and $R_{z,signal}$ (right) in the center of the active layer. We can notice that both pump and signal wavelengths are well guided all along the waveguide.

\subsection{Population map}
\begin{figure}[h!]
	\centering
	\includegraphics[width=0.49\textwidth,trim=0cm 0cm 0cm 1cm,clip]{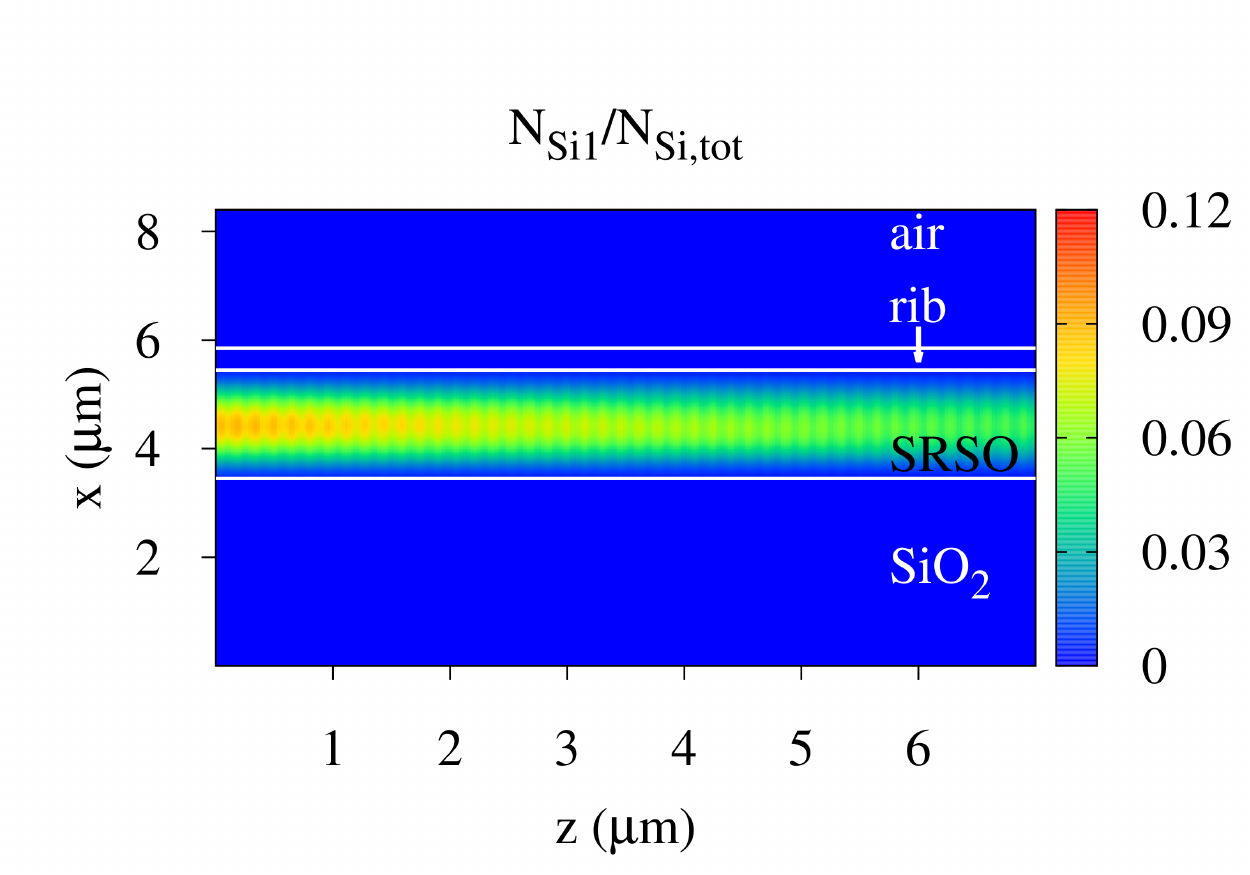}
	\includegraphics[width=0.49\textwidth,trim=0cm 0cm 0cm 1cm,clip]{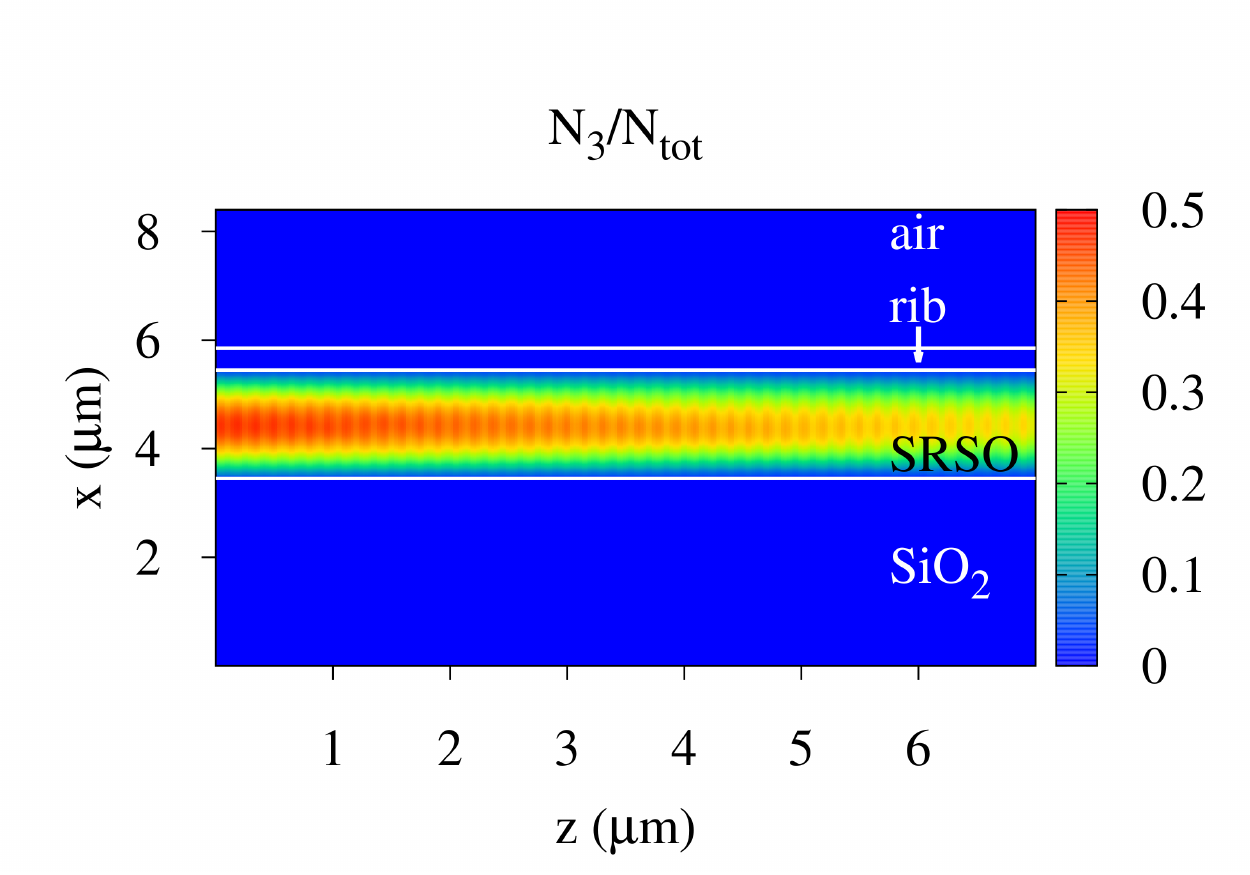}
	\caption{Longitudinal section view of $ N_{Si}^*/N_{Si_{tot}} $ (on the left) and $N_3/N_{tot}$ (on the right)}
	\label{populationmap}
\end{figure}

The ADE-FDTD method allows to compute tridimensional population distributions in different states. We present the population ratio distributions in the waveguide in a longitudinal section view along the propagation axis in the Fig. \ref{populationmap}. On the left, the ratio between the excited level population ($ N_{Si_1} $) and the total number of Si-ng ($N_{Si}= N_{Si_0} + N_{Si_1}  $) and, on the right, the ratio between the level population $ N_3 $ and the total number of $ \mathrm{Nd^{3+}} $ ions for a pump power equal to $ 10^3~\mathrm{mW.mm^{-2}}$. The maximum percentage of the excited Si-ng is 10\%, and the maximum percentage of the excited $ Nd^{3+} $ is 50\%. With a transfer coefficient taken K is equal to 0, the percentage of the excited $ Nd^{3+} $ is close to 0\%. This confirms the major role of energy transfer in the exciting process of neodymium ions.

The decreasing profile $N_{Si_1}(z)$ of the excited Si-ng is consistent with the pump profile $R_{z,pump}(z)$ decrease in Fig. \ref{fieldmap} due to the absorption of the pump field by the Si-ng. Hence, the excited level population $\mathrm{N_3}$ decreases by 30 \% over the 7 $\mathrm{\mu m}$ length of the waveguide. This imply that co-propagation pumping scheme is not a good method in order to reach a uniformly pump longer waveguide. A top pumping configuration along the waveguide length would probably lead to a more uniformly pumped active layer but is not the purpose of the present paper. 

\subsection{Calculation of the gain}
From population distributions, we calculate the gross gain of the transition of our interest $ N_3 \rightarrow N_1 (^4 F_{3/2}\rightarrow ^4 I_{11/2},\:\lambda_{31}=1064nm)$. The local gross gain per unit length at 1064 nm is given by: 
\begin{equation}
g_{\mathrm{dB.cm^{-1}}}(x,y,z)= \frac{10}{\ln 10}\left(\sigma_{em} N_3(x,y,z)- \sigma_{abs} N_1(x,y,z))\right)
\label{eq_gain}
\end{equation}

where $ \sigma_{abs} $ and $ \sigma_{em} $ are respectively the absorption and the emission cross sections. Due to the short decay lifetime ($\tau_{10}$)  from level 1 to the ground level 0, we observe that: $ \mathrm{N_1}\ll \mathrm{N_3} $. Moreover, we consider that $ \sigma_{em} $ is comparable to $ \sigma_{abs} $ thus Eq. (\ref{eq_gain}) becomes:
\begin{equation}
g_{\mathrm{dB.cm^{-1}}}(x,y,z) \simeq \frac{10}{\ln 10}\left(\sigma_{em} N_3(x,y,z)\right)
\end{equation}

\begin{figure}[h!]
	\centering
	\includegraphics[width=0.8\textwidth,trim=1cm 1cm 1cm 1cm]{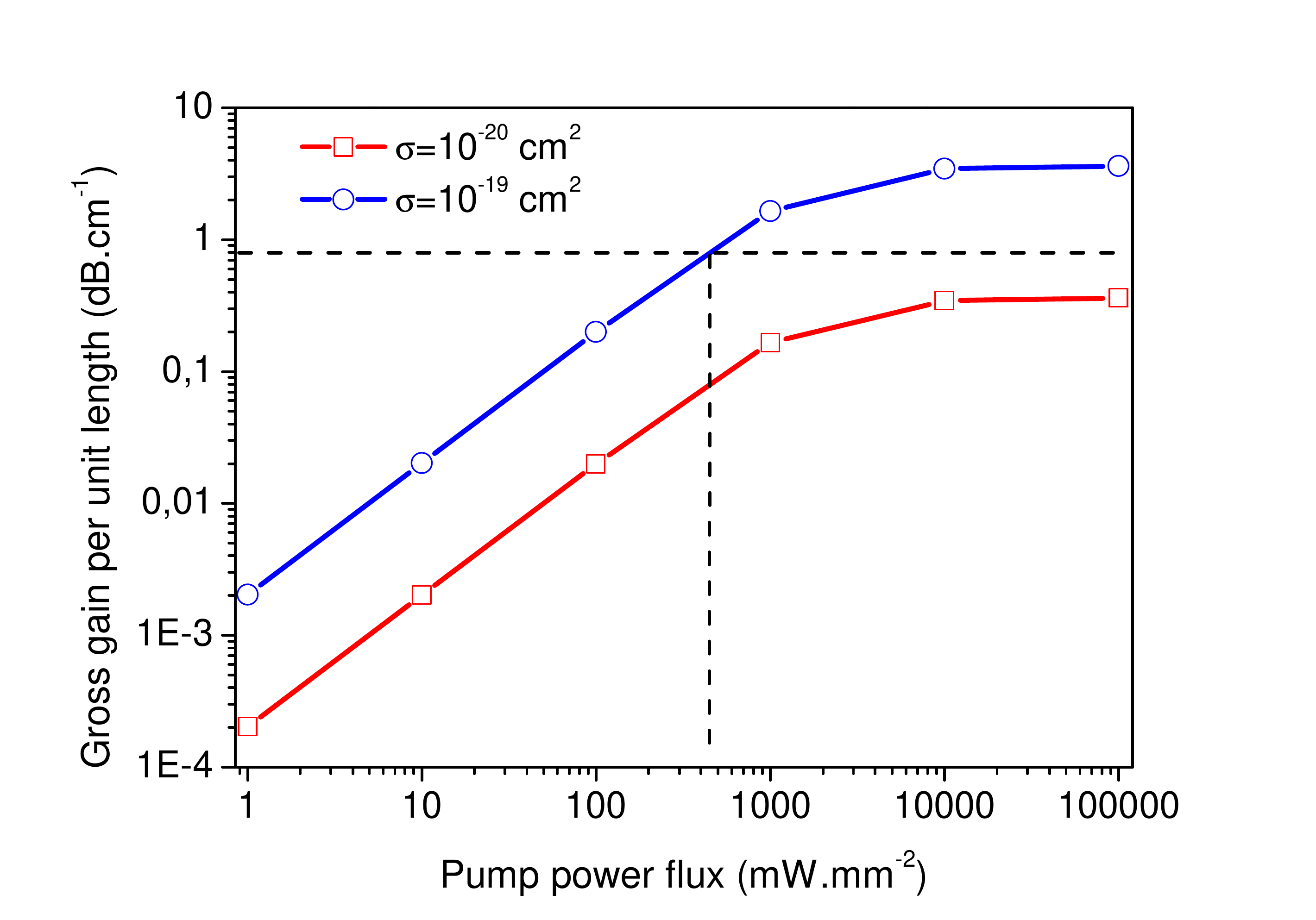}
	\caption{Local gross gain per unit length at the center of the active layer as a function of the pumping power, (horizontal dashed line) Losses of 0.8 $\mathrm{dB.cm^{-1}}$ found by Pirastesh et al\cite{Pirasteh2012}.}	\label{gain}
\end{figure}

For two extreme values of the emission cross section ($\sigma_{em}=10^{-20}~\mathrm{cm^{2}}$ and $\sigma_{em}=10^{-19}~\mathrm{cm^{2}}$), we calculate the local gross gain per unit length versus different values of pump intensity in Fig. \ref{gain} at the point $(x_c,y_c,z_c)$, where $ x_c,y_c $ are the coordinates of the center of the active layer and $ z_c $ is the middle of the waveguide. We can notice that the local gross gain per unit length saturates for pumping powers higher than $10^{5}~\mathrm{mW.mm^{-2}}$ and reach 0.36$~\mathrm{dB.cm^{-1}}$ for $\sigma_{em}=10^{-20}~\mathrm{cm^{2}}$ and  3.6$~\mathrm{dB.cm^{-1}}$ for $\sigma_{em}=10^{-19}~\mathrm{cm^{2}}$ respectively. In a similar waveguide, Pirasteh et al\cite{Pirasteh2012} determined experimentally losses equal to $\alpha=0.8~\mathrm{dB.cm^{-1}}$, represented by a horizontal dashed line on Fig. \ref{gain}. For the lower emission cross section $\sigma_{em}=10^{-20}~cm^2 $ the local gross gain cannot compensate the experimentally determined losses of $0.8~\mathrm{dB~cm^{-1}}$. For the highest emission cross section $\sigma_{em}=10^{-19}~\mathrm{cm^{2}}$, it is necessary to pump the waveguide with power higher than a threshold value of $450~\mathrm{mW.mm^{-2}}$ to obtain internal net gain. By linear extrapolation and with a high pump power equal to $10^{5}~\mathrm{mW.mm^{-2}}$ we can estimate a threshold value of cross section $\sigma_{em}= 2.2\times10^{-20}~\mathrm{cm^{2}}$ at which positive internal net gain is reached. Despite, more accurate measurement of the emission cross section will lead to better estimation of the possible internal net gain with such a waveguide. This study shows that to increase the gain, several ways may be explore: (i) An increase of the neodymium and Si-ng concentration may lead to higher gain, however some limits to concentration may occurs above some $10^{20}~\mathrm{dopants.cm^{-3}}$(ii) An increase of the coupling efficiency or of the fraction of excited rare earth may lead to higher gain (iii) A top pumping configuration  may result in a more uniformly pumped active layer and consequently in more uniform distribution of gain along the waveguide length. These will be the object of further experimental and theoretical studies.

\section{Conclusion}
A new algorithm based on ADE-FDTD method has been presented that allow to describe the spatial distribution of the eletromagnetic field and levels population in steady state in an active optical waveguide. The multi-scale times issue of such a system has been overcome by the development of this algorithm leading a drastic reduction of number of iterations to reach steady states values of fields and levels population (from $10^{15}$  to $10^{5}$ iterations). Moreover, we proposed a method to calibrate the $i\rightarrow j$ transition linewidth $ \Delta \omega_{ij}$ according to the experimental absorption cross section, making a possible comparison between experimental and theoretical studies. We apply our new algorithm to a strip loaded waveguide whose active layer is constituted of a silicon rich silicon oxide (SRSO) layer doped with silicon nanograin and neodymium ions. Using physical parameters such as absorption cross section ranging from $10^{-20}~\mathrm{cm^2}$ to $10^{-19}~\mathrm{cm^2}$ and concentrations in accordance with the literature, we found a gross gain ranging from 0.36$\mathrm{~dB.cm^{-1}}$ to 3.6$\mathrm{~dB.cm^{-1}}$, based on experimental losses we found a threshold pump power value of 450$~\mathrm{mW.mm^{-2}}$ necessary to have a positive net gain. We would like to emphasize the point that the method developed here is generalizable to other systems presenting very different characteristic times resulting in a drastic reduction of the calculation time for reaching steady states.

\bigskip

\section*{Acknowledgments}
The authors are grateful to the French Nation Research Agency, which supported this work through the Nanoscience and Nanotechnology program (DAPHNES project ANR-08-NANO-005).

\begin{thebibliography}{10}
\newcommand{\enquote}[1]{``#1''}

\bibitem{Yee1966}
K.~Yee, \enquote{Numerical solution of initial boundary value problems
  involving maxwell's equations in isotropic media,}
  IEEE Trans. Antennas Propag. \textbf{14}, 302--307 (1966).

\bibitem{Taflove1995}
A.~Taflove and S.~C. Hagness, \emph{Computational Electrodynamics: the
  Finite-Difference Time-Domain Method} (Artech House, 1995).

\bibitem{Miniscalco1991}
W.~Miniscalco, \enquote{Erbium-doped glasses for fiber amplifiers at 1500 nm,}
  J. Lightwave Techno. \textbf{9}, 234 --250 (1991).

\bibitem{kik1998erbium}
P.~Kik and A.~Polman, \enquote{Erbium-doped optical-waveguide amplifiers on
  silicon,} Mater. Res. Bull. \textbf{23}, 48--54 (1998).

\bibitem{Polman2004}
A.~Polman and F.~C. J.~M. van Veggel, \enquote{Broadband sensitizers for
  erbium-doped planar optical amplifiers: review,} J. Opt. Soc. Am. B
  \textbf{21}, 871--892 (2004).

\bibitem{Kenyon1994}
A.~J. Kenyon, P.~F. Trwoga, M.~Federighi, and C.~W. Pitt, \enquote{Optical
  properties of {PECVD} erbium-doped silicon-rich silica: evidence for energy
  transfer between silicon microclusters and erbium ions,} J. Phys.:
  Condens. Matter \textbf{6}, 319--324 (1994).

\bibitem{Fujii1997}
M.~Fujii, M.~Yoshida, Y.~Kanzawa, S.~Hayashi, and K.~Yamamoto, \enquote{1.54
  µm photoluminescence of {E}r3+ doped into {S}i{O}2 films containing {S}i
  nanocrystals: {E}vidence for energy transfer from {S}i nanocrystals to
  {E}r3+,} Appl. Phys. Lett. \textbf{71}, 1198 --1200 (1997).

\bibitem{Dufour2011}
C.~Dufour, J.~Cardin, O.~Debieu, A.~Fafin, and F.~Gourbilleau,
  \enquote{Electromagnetic modeling of waveguide amplifier based on {N}d3+
  {S}i-rich {S}i{O}2 layers by means of the {ADE-FDTD} method,} Nanoscale  Res. Lett. \textbf{6}, 1--5 (2011).

\bibitem{hagness1996subpicosecond}
S.~C. Hagness, R.~M. Joseph, and A.~Taflove, \enquote{Subpicosecond
  electrodynamics of distributed {B}ragg reflector microlasers: {R}esults from
  finite difference time domain simulations,} Radio Sci. \textbf{31},
  931--941 (1996).

\bibitem{Berenger1994}
J.~Berenger, \enquote{A perfectly matched layer for the absorption of
  electromagnetic waves,} J. comput. phys. \textbf{114},
  185--200 (1994).

\bibitem{taflove1975numerical}
A.~Taflove and M.~E. Brodwin, \enquote{Numerical solution of steady-state  electromagnetic scattering problems using the time-dependent {M}axwell's equations,} IEEE Trans. Microwave Theory Tech. \textbf{23},
  623--630 (1975).

\bibitem{chang2004finite}
S.-H. Chang and A.~Taflove, \enquote{Finite-difference time-domain model of
  lasing action in a four-level two-electron atomic system,} Opt. Express  \textbf{12}, 3827--3833 (2004).

\bibitem{Siegman1986}
A.~E. Siegman, \emph{Lasers} (University Science Books, 1986).

\bibitem{Petropoulos1994}
P.G.~Petropoulos, \enquote{Stability and phase error analysis of {FD-TD} in dispersive dielectrics,} IEEE Trans. Antennas Propag. \textbf{42}, 62 --69 (1994).

\bibitem{Nagra1998}
A.~Nagra and R.~York, \enquote{{FDTD} analysis of wave propagation in nonlinear absorbing and gain media,} IEEE Trans. Antennas Propag. \textbf{46}, 334 --340 (1998).

\bibitem{Pacifici2003}
D.~Pacifici, G.~Franzo, F.~Priolo, F.~Iacona, and L.~Dal~Negro,
  \enquote{Modeling and perspectives of the {S}i nanocrystals-{E}r interaction for optical amplification,} Phys. Rev. B \textbf{67}, 245301, 1--13 (2003).

\bibitem{Lee2005}
H.~Lee, J.~Shin, and N.~Park, \enquote{Performance analysis of nanocluster-{S}i sensitized {E}r doped waveguide amplifier using top-pumped 470nm {LED},} Opt. Express \textbf{13}, 9881--9889 (2005).

\bibitem{Toccafondo2008}
V.~Toccafondo, S.~Faralli, and F.~Di~Pasquale, \enquote{Evanescent {M}ultimode {L}ongitudinal {P}umping {S}cheme for {S}i-{N}anocluster {S}ensitized {E}r$^{3+}$ {D}oped {W}aveguide {A}mplifiers,} J. Lightwave Techno. \textbf{26}, 3584--3591 (2008).

\bibitem{Oubre2004}
C.~Oubre and P.~Nordlander, \enquote{Optical properties of metallodielectric nanostructures calculated using the finite difference time domain method,} J. Phys. Chem. B \textbf{108}, 17740--17747 (2004).

\bibitem{Biallo2006}
D.~Biallo, A.~D'Orazio, and V.~Petruzzelli, \enquote{Enhanced light extraction in {E}r$^{3+}$ doped {S}i{O}2-{T}i{O}2 microcavity embedded in   one-dimensional photonic crystal,} J. Non-Cryst. Solids \textbf{352}, 3823--3828 (2006).

\bibitem{Fallahkhair2008}
A.~Fallahkhair, K.~Li, and T.~Murphy, \enquote{Vector finite difference
  modesolver for anisotropic dielectric waveguides,} J. Lightwave Techno. \textbf{26}, 1423--1431 (2008).

\bibitem{MUMPS:1}
P.~R. Amestoy, I.~S. Duff, J.~Koster, and J.-Y. L'Excellent, \enquote{A {F}ully {A}synchronous {M}ultifrontal {S}olver {U}sing {D}istributed {D}ynamic {S}cheduling,} SIAM. J. Matrix Anal. \& Appl. \textbf{23},
  15--41 (2001).

\bibitem{MUMPS:2}
P.~R. Amestoy, A.~Guermouche, J.-Y. L'Excellent, and S.~Pralet, \enquote{Hybrid scheduling for the parallel solution of linear systems,} Parallel Computing \textbf{32}, 136--156 (2006).

\bibitem{Debieu2011}
O.~Debieu, J.~Cardin, X.~Portier, and F.~Gourbilleau, \enquote{Effect of the {N}d content on the structural and photoluminescence properties of   silicon-rich silicon dioxide thin films,} Nanoscale Res. Lett.   \textbf{6}, 1--8 (2011).

\bibitem{Govoni2012}
M.~Govoni, .~Marri, and S.~Ossicini, \enquote{Carrier multiplication between interacting nanocrystals for fostering silicon-based photovoltaics,} Nat. Photonics \textbf{6}, 672--679 (2012).

\bibitem{Priolo2001}
F.~Priolo, G.~Franzo, D.~Pacifici, V.~Vinciguerra, F.~Iacona, and A.~Irrera, \enquote{Role of the energy transfer in the optical properties of undoped and {E}r-doped interacting {S}i nanocrystals,} J. Appl. Phys. \textbf{89}, 264--272 (2001).

\bibitem{Pirasteh2012}
P.~Pirasteh, J.~Charrier, Y.~Dumeige, Y.~G. Boucher, O.~Debieu, and
  F.~Gourbilleau, \enquote{Study of optical losses of {N}d$^{3+}$ doped silicon rich silicon oxide for laser cavity,} Thin Solid Films \textbf{520}, 4026--4030 (2012).

\end{thebibliography}
\end{document}